\PassOptionsToPackage{unicode}{hyperref}
\PassOptionsToPackage{hyphens}{url}
\documentclass[
  11pt,
]{article}
\usepackage{xcolor}
\usepackage[margin=1in]{geometry}
\usepackage{amsmath,amssymb}
\usepackage{iftex}
\ifPDFTeX
  \usepackage[T1]{fontenc}
  \usepackage[utf8]{inputenc}
  \usepackage{textcomp} 
\else 
  \usepackage{unicode-math} 
  \defaultfontfeatures{Scale=MatchLowercase}
  \defaultfontfeatures[\rmfamily]{Ligatures=TeX,Scale=1}
\fi
\usepackage{lmodern}
\ifPDFTeX\else
\fi
\IfFileExists{upquote.sty}{\usepackage{upquote}}{}
\IfFileExists{microtype.sty}{
  \usepackage[]{microtype}
  \UseMicrotypeSet[protrusion]{basicmath} 
}{}
\makeatletter
\@ifundefined{KOMAClassName}{
  \IfFileExists{parskip.sty}{%
    \usepackage{parskip}
  }{
    \setlength{\parindent}{0pt}
    \setlength{\parskip}{6pt plus 2pt minus 1pt}}
}{
  \KOMAoptions{parskip=half}}
\makeatother
\usepackage{color}
\usepackage{fancyvrb}

\DefineVerbatimEnvironment{Highlighting}{Verbatim}{commandchars=\\\{\}}
\usepackage{framed}
\definecolor{shadecolor}{RGB}{248,248,248}
\newenvironment{Shaded}{\begin{snugshade}}{\end{snugshade}}

\newcommand{\AttributeTok}[1]{\textcolor[rgb]{0.13,0.29,0.53}{#1}}

\newcommand{\CommentTok}[1]{\textcolor[rgb]{0.56,0.35,0.01}{\textit{#1}}}

\newcommand{\DecValTok}[1]{\textcolor[rgb]{0.00,0.00,0.81}{#1}}

\newcommand{\FloatTok}[1]{\textcolor[rgb]{0.00,0.00,0.81}{#1}}
\newcommand{\FunctionTok}[1]{\textcolor[rgb]{0.13,0.29,0.53}{\textbf{#1}}}

\newcommand{\NormalTok}[1]{#1}

\newcommand{\OtherTok}[1]{\textcolor[rgb]{0.56,0.35,0.01}{#1}}

\newcommand{\SpecialCharTok}[1]{\textcolor[rgb]{0.81,0.36,0.00}{\textbf{#1}}}

\newcommand{\StringTok}[1]{\textcolor[rgb]{0.31,0.60,0.02}{#1}}

\usepackage{graphicx}
\makeatletter
\newsavebox\pandoc@box
\newcommand*\pandocbounded[1]{
  \sbox\pandoc@box{#1}%
  \Gscale@div\@tempa{\textheight}{\dimexpr\ht\pandoc@box+\dp\pandoc@box\relax}%
  \Gscale@div\@tempb{\linewidth}{\wd\pandoc@box}%
  \ifdim\@tempb\p@<\@tempa\p@\let\@tempa\@tempb\fi
  \ifdim\@tempa\p@<\p@\scalebox{\@tempa}{\usebox\pandoc@box}%
  \else\usebox{\pandoc@box}%
  \fi%
}
\def\fps@figure{htbp}
\makeatother
\NewDocumentCommand\citeproctext{}{}
\NewDocumentCommand\citeproc{mm}{%
  \begingroup\def\citeproctext{#2}\cite{#1}\endgroup}
\makeatletter
 \let\@cite@ofmt\@firstofone
 \def\@biblabel#1{}
 \def\@cite#1#2{{#1\if@tempswa , #2\fi}}
\makeatother
\newlength{\cslhangindent}
\newlength{\csllabelwidth}
\newenvironment{CSLReferences}[2] 
 {\begin{list}{}{%
  \setlength{\itemindent}{0pt}
  \setlength{\leftmargin}{0pt}
  \setlength{\parsep}{0pt}
  \ifodd #1
   \setlength{\leftmargin}{\cslhangindent}
   \setlength{\itemindent}{-1\cslhangindent}
  \fi
  \setlength{\itemsep}{#2\baselineskip}}}
 {\end{list}}
\usepackage{calc}

\providecommand{\tightlist}{%
  \setlength{\itemsep}{0pt}\setlength{\parskip}{0pt}}
\usepackage{hyperref}
\usepackage{amsmath}
\usepackage{float}

\usepackage{booktabs}
\usepackage{longtable}
\usepackage{array}
\usepackage{multirow}
\usepackage{wrapfig}
\usepackage{float}
\usepackage{colortbl}
\usepackage{pdflscape}
\usepackage{tabu}
\usepackage{threeparttable}
\usepackage{threeparttablex}
\usepackage[normalem]{ulem}
\usepackage{makecell}
\usepackage{xcolor}
\usepackage{bookmark}
\IfFileExists{xurl.sty}{\usepackage{xurl}}{} 
\makeatletter
\@ifundefined{xmpquote}{}{}
\makeatother
\hypersetup{
  pdftitle={bayprior: Structured Bayesian Prior Elicitation{,} Conflict Diagnostics{,} and Regulatory Reporting},
  hidelinks,
  pdfcreator={LaTeX via pandoc}}

\title{bayprior: Structured Bayesian Prior Elicitation, Conflict
Diagnostics, and Regulatory Reporting}
\author{Ndoh Penn\\
Independent Researcher\\
Reprostats, Belgium}
\date{2026-10-07}

\begin{document}
\maketitle
\begin{abstract}
Bayesian clinical trials require a well-specified prior distribution,
and regulators increasingly expect that specification to be documented,
diagnosed for prior-data conflict, and shown to be robust to reasonable
alternatives, yet few R tools connect the full workflow of prior
construction, validation, and regulatory justification into a single
package. We introduce bayprior, an R package and Shiny application that
implements structured expert elicitation via quantile matching, moment
matching, and the SHELF roulette method across six distribution
families; linear and logarithmic expert opinion pooling with
Bhattacharyya agreement diagnostics; derivation of
meta-analytic-predictive (MAP) priors from historical trial data via
random-effects meta-analysis; prior-data conflict assessment using the
Box p-value, S-value (surprisal), surprise index, information
divergence, Bhattacharyya overlap, and multivariate Mahalanobis
distance; hyperparameter sensitivity analysis with tornado and influence
heatmap visualisations; and robust, sceptical, and power prior
alternatives. A regulatory reporting module generates self-contained
HTML, PDF, or Word documents addressing expectations in the FDA's 2026
draft Bayesian methods guidance and in EMA guidance on incorporating
external and historical information. The package is demonstrated on a
synthetic oncology Phase II trial. bayprior is available on CRAN and
includes a fully modular Shiny application for interactive use.
\end{abstract}

\section{Introduction}\label{introduction}

Bayesian statistical methods are increasingly used in clinical trial
design and analysis, offering flexible incorporation of prior knowledge
and natural uncertainty quantification (\citeproc{ref-FDA2026}{U.S. Food
and Drug Administration 2026}). However, the widespread adoption of
Bayesian methods in regulatory settings requires that prior
distributions be not only well-chosen but also transparently documented
and demonstrably robust. The FDA's 2026 draft guidance on Bayesian
methods (\citeproc{ref-FDA2026}{U.S. Food and Drug Administration 2026})
calls for sponsors to demonstrate that trial conclusions hold under
sensitivity analyses over the prior specification, but does not
prescribe specific software tools or diagnostics for doing so.

In practice, prior specification often proceeds informally. Analysts may
elicit a prior from clinical literature or expert opinion without a
structured procedure, assess prior-data conflict informally by visual
inspection, and document the justification in narrative form. This
approach is difficult to reproduce and rarely satisfies regulatory
reviewers who expect a systematic, auditable justification.

Existing R packages for Bayesian clinical trials focus on posterior
computation and adaptive design. \texttt{RBesT}
(\citeproc{ref-RBesT}{Weber et al. 2025}) implements robust
meta-analytic-predictive priors for historical data borrowing.
\texttt{trialr} (\citeproc{ref-trialr}{Brock 2025}) provides a range of
adaptive trial designs via Stan. \texttt{hdbayes}
(\citeproc{ref-hdbayes}{Alt et al. 2025}) implements multiple historical
data borrowing priors (including power priors, normalised power priors,
robust MAP priors, and commensurate priors) for generalised linear
models. Each of these tools addresses a specific stage of the Bayesian
workflow well, but none carries a prior through construction from expert
knowledge, prior-data conflict assessment, and regulatory reporting as a
single connected process.

We introduce \texttt{bayprior}, an R package and Shiny application that
connects these stages by providing an integrated prior justification
workflow: structured expert elicitation, expert opinion pooling,
prior-data conflict diagnostics, sensitivity analysis, robust and
regulatory prior alternatives, and automated regulatory reporting. The
package implements established statistical methodology
(\citeproc{ref-OHagan2006}{O'Hagan et al. 2006};
\citeproc{ref-Box1980}{Box 1980}; \citeproc{ref-Oakley2010}{Oakley and
O'Hagan 2010}; \citeproc{ref-Schmidli2014}{Schmidli et al. 2014};
\citeproc{ref-Ibrahim2000}{Ibrahim and Chen 2000};
\citeproc{ref-Spiegelhalter1994}{Spiegelhalter et al. 1994}) in a
unified, accessible framework suitable for both programmatic use and
interactive exploration via the bundled Shiny application.

\section{Related software}\label{related-software}

\subsection{Expert Elicitation}\label{expert-elicitation}

The Sheffield Elicitation Framework (\citeproc{ref-Oakley2010}{Oakley
and O'Hagan 2010}) provides a structured approach to expert elicitation
and is implemented in the R package \texttt{SHELF}
(\citeproc{ref-SHELF}{Oakley 2026}). \texttt{SHELF} supports elicitation
from individual or multiple experts across a range of distribution
families, with weighted linear pooling for aggregating multiple expert
beliefs. \texttt{bayprior} extends this with three programmatic
elicitation methods (quantile matching, moment matching, and the
roulette method) across six distribution families, adds logarithmic
opinion pooling as a complement to linear pooling, and introduces
quantitative agreement diagnostics via the Bhattacharyya coefficient for
assessing inter-expert consistency before pooling.

\subsection{Robust Priors and Historical
Borrowing}\label{robust-priors-and-historical-borrowing}

\texttt{RBesT} (\citeproc{ref-RBesT}{Weber et al. 2025}) implements the
robust meta-analytic-predictive (MAP) prior approach
(\citeproc{ref-Schmidli2014}{Schmidli et al. 2014}) for incorporating
historical control data. \texttt{hdbayes} (\citeproc{ref-hdbayes}{Alt et
al. 2025}) provides a broader range of historical borrowing methods,
including power priors, normalised power priors, commensurate priors,
and robust MAP priors, for generalised linear models via Stan.
\texttt{bayprior} implements the same MAP approach
(\citeproc{ref-Schmidli2014}{Schmidli et al. 2014};
\citeproc{ref-Neuenschwander2010}{Neuenschwander et al. 2010}) as a
closed-form random-effects meta-analysis with an explicit,
outcome-type-specific prior on the between-trial heterogeneity parameter
\(\tau\) (\citeproc{ref-Roever2021}{Roever et al. 2021}), requiring
neither Stan nor an external sampler, and provides the Schmidli robust
mixture prior and calibrated power priors
(\citeproc{ref-Ibrahim2000}{Ibrahim and Chen 2000}) as further
alternatives within the same prior justification workflow.

\subsection{Prior Sensitivity and Conflict
Diagnostics}\label{prior-sensitivity-and-conflict-diagnostics}

\texttt{priorsense} (\citeproc{ref-priorsense}{Kallioinen et al. 2025})
implements power-scaling sensitivity analysis for detecting prior-data
conflict and likelihood noninformativity in fitted Stan models
(\citeproc{ref-Kallioinen2023}{Kallioinen et al. 2023}).
\texttt{bayprior} addresses a complementary use case: closed-form
conflict diagnostics for conjugate prior families at the prior
specification stage, before any model is fitted, without requiring Stan.
The two packages serve different points in the Bayesian workflow.

\subsection{Scope Comparison}\label{scope-comparison}

Table 1 summarises the scope of existing packages relative to
\texttt{bayprior}.

\begin{longtable}[t]{llllllll}
\caption{\label{tab:pkg-comparison}Scope comparison of R packages for Bayesian prior specification.
             Partial: SHELF pooling supports linear pooling only.
             Hist. borrowing: derives an informative prior from historical
             trial-level data (as opposed to Robust priors, which modifies
             an already-specified prior). priorsense conflict and
             sensitivity diagnostics require a fitted Stan model.}\\
\toprule
Package & Elicitation & Pooling & Hist. borrowing & Conflict & Sensitivity & Robust priors & Reporting\\
\midrule
SHELF & Yes & Partial & No & No & No & No & No\\
RBesT & No & No & Yes & No & No & Yes & No\\
trialr & No & No & No & No & No & No & No\\
hdbayes & No & No & Yes & No & No & Yes & No\\
priorsense & No & No & No & Yes & Yes & No & No\\
\addlinespace
bayprior & Yes & Yes & Yes & Yes & Yes & Yes & Yes\\
\bottomrule
\end{longtable}

\section{Package overview}\label{package-overview}

\texttt{bayprior} is built using the \texttt{golem}
(\citeproc{ref-golem}{Fay et al. 2024}) framework and provides both
exported R functions for programmatic use and a modular Shiny
application for interactive exploration. The package is available on
CRAN and can be installed with:

\begin{Shaded}
\begin{Highlighting}[]
\FunctionTok{install.packages}\NormalTok{(}\StringTok{"bayprior"}\NormalTok{)}

\CommentTok{\# Or the development version from GitHub:}
\NormalTok{pak}\SpecialCharTok{::}\FunctionTok{pak}\NormalTok{(}\StringTok{"ndohpenngit/bayprior"}\NormalTok{)}
\end{Highlighting}
\end{Shaded}

The Shiny application can be launched with:

\begin{Shaded}
\begin{Highlighting}[]
\FunctionTok{library}\NormalTok{(bayprior)}
\FunctionTok{run\_app}\NormalTok{()}
\end{Highlighting}
\end{Shaded}

\subsection{Prior Elicitation}\label{prior-elicitation}

\texttt{bayprior} implements three structured elicitation approaches
across six distribution families (Beta, Normal, Gamma, Log-Normal,
Exponential, and Weibull).

\textbf{Moment matching} derives hyperparameters analytically from an
expert-supplied mean and standard deviation:

\begin{Shaded}
\begin{Highlighting}[]
\NormalTok{prior }\OtherTok{\textless{}{-}} \FunctionTok{elicit\_beta}\NormalTok{(}
  \AttributeTok{mean      =} \FloatTok{0.35}\NormalTok{,}
  \AttributeTok{sd        =} \FloatTok{0.10}\NormalTok{,}
  \AttributeTok{method    =} \StringTok{"moments"}\NormalTok{,}
  \AttributeTok{label     =} \StringTok{"Response rate"}\NormalTok{,}
  \AttributeTok{expert\_id =} \StringTok{"Expert\_1"}
\NormalTok{)}
\end{Highlighting}
\end{Shaded}

\begin{longtable}[t]{ll}
\caption{\label{tab:prior-table}Elicited Beta prior for the response rate (moment matching).}\\
\toprule
Quantity & Value\\
\midrule
Distribution & BETA\\
alpha & 7.6125\\
beta & 14.1375\\
Method & moments elicitation\\
Expert & Expert\_1\\
\addlinespace
Mean (SD) & 0.35 (0.1)\\
95\% CrI & {}[0.1694, 0.5567]\\
\bottomrule
\end{longtable}

\textbf{Quantile matching} fits the distribution numerically to
expert-specified probability-value pairs, accommodating asymmetric
beliefs that moment matching cannot capture. For example, an expert who
believes the response rate has a 25\% chance of being below 0.20, a 60\%
chance of being below 0.35, and an 80\% chance of being below 0.50:

\begin{Shaded}
\begin{Highlighting}[]
\NormalTok{prior\_qm }\OtherTok{\textless{}{-}} \FunctionTok{elicit\_beta}\NormalTok{(}
  \AttributeTok{quantiles =} \FunctionTok{c}\NormalTok{(}\StringTok{\textasciigrave{}}\AttributeTok{0.25}\StringTok{\textasciigrave{}} \OtherTok{=} \FloatTok{0.20}\NormalTok{, }\StringTok{\textasciigrave{}}\AttributeTok{0.60}\StringTok{\textasciigrave{}} \OtherTok{=} \FloatTok{0.35}\NormalTok{, }\StringTok{\textasciigrave{}}\AttributeTok{0.80}\StringTok{\textasciigrave{}} \OtherTok{=} \FloatTok{0.50}\NormalTok{),}
  \AttributeTok{method    =} \StringTok{"quantile"}\NormalTok{,}
  \AttributeTok{label     =} \StringTok{"Response rate (quantile)"}\NormalTok{,}
  \AttributeTok{expert\_id =} \StringTok{"Expert\_1"}
\NormalTok{)}
\end{Highlighting}
\end{Shaded}

\begin{Shaded}
\begin{Highlighting}[]
\FunctionTok{plot}\NormalTok{(prior)}
\end{Highlighting}
\end{Shaded}

\begin{figure}[H]
\includegraphics[width=1\linewidth]{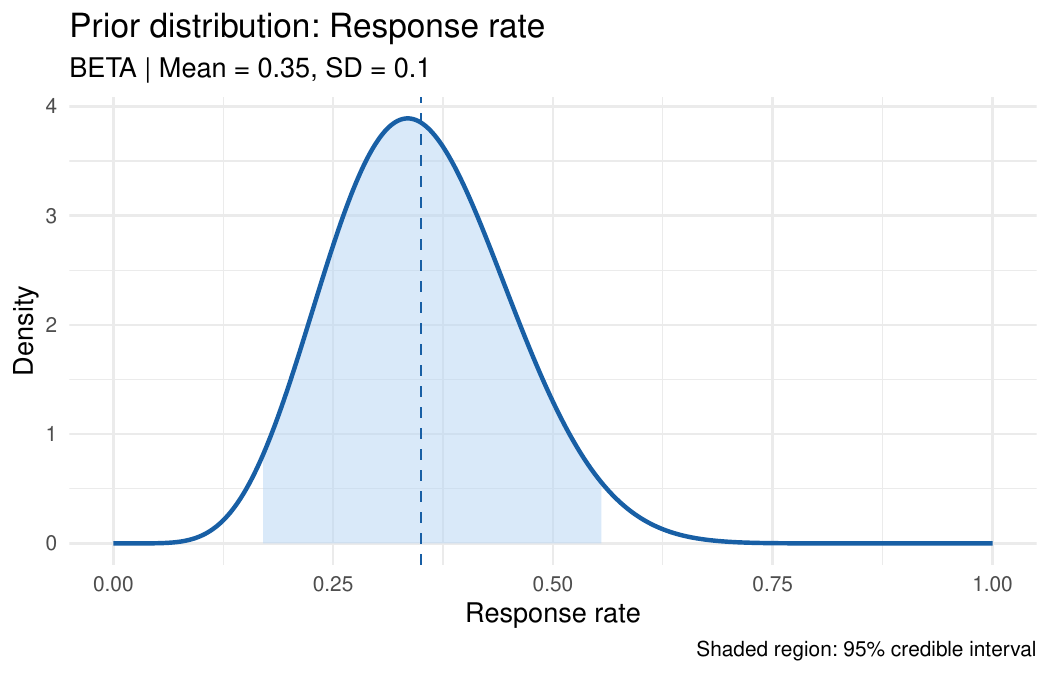} \caption{Beta prior elicited by moment matching (mean = 0.35, SD = 0.10). Shaded region shows the 95\% credible interval.}\label{fig:elicitation-plot}
\end{figure}

The fitted distribution, parameter estimates, and a 95\% credible
interval are displayed automatically. All six families share a common
API, so \texttt{elicit\_beta()} can be replaced with
\texttt{elicit\_normal()}, \texttt{elicit\_gamma()},
\texttt{elicit\_lognormal()}, \texttt{elicit\_exponential()}, or
\texttt{elicit\_weibull()} with identical syntax.

\subsection{Expert Pooling}\label{expert-pooling}

Multiple expert priors can be aggregated using linear or logarithmic
opinion pooling (\citeproc{ref-OHagan2006}{O'Hagan et al. 2006}). Linear
pooling takes the weighted mixture of the densities,
\(\pi(\theta) = \sum_k w_k \pi_k(\theta)\); logarithmic pooling takes
the weighted geometric mean, \(\pi(\theta) \propto
\prod_k \pi_k(\theta)^{w_k}\). For linear pooling, the consensus mean
and SD are computed exactly from the component means/SDs and weights,
without numerical integration:
\[\text{mean} = \sum_k w_k \bar{x}_k, \qquad
  \text{var} = \sum_k w_k \left(s_k^2 + \bar{x}_k^2\right) - \text{mean}^2\]
No closed-form SD exists for logarithmic pooling in general.

\begin{Shaded}
\begin{Highlighting}[]
\NormalTok{e1 }\OtherTok{\textless{}{-}} \FunctionTok{elicit\_beta}\NormalTok{(}\AttributeTok{mean =} \FloatTok{0.30}\NormalTok{, }\AttributeTok{sd =} \FloatTok{0.08}\NormalTok{, }\AttributeTok{method =} \StringTok{"moments"}\NormalTok{,}
                  \AttributeTok{label =} \StringTok{"Response rate"}\NormalTok{, }\AttributeTok{expert\_id =} \StringTok{"Expert\_1"}\NormalTok{)}
\NormalTok{e2 }\OtherTok{\textless{}{-}} \FunctionTok{elicit\_beta}\NormalTok{(}\AttributeTok{mean =} \FloatTok{0.42}\NormalTok{, }\AttributeTok{sd =} \FloatTok{0.10}\NormalTok{, }\AttributeTok{method =} \StringTok{"moments"}\NormalTok{,}
                  \AttributeTok{label =} \StringTok{"Response rate"}\NormalTok{, }\AttributeTok{expert\_id =} \StringTok{"Expert\_2"}\NormalTok{)}

\NormalTok{consensus }\OtherTok{\textless{}{-}} \FunctionTok{aggregate\_experts}\NormalTok{(}
  \AttributeTok{priors  =} \FunctionTok{list}\NormalTok{(}\AttributeTok{E1 =}\NormalTok{ e1, }\AttributeTok{E2 =}\NormalTok{ e2),}
  \AttributeTok{weights =} \FunctionTok{c}\NormalTok{(}\FloatTok{0.5}\NormalTok{, }\FloatTok{0.5}\NormalTok{),}
  \AttributeTok{method  =} \StringTok{"linear"}
\NormalTok{)}
\end{Highlighting}
\end{Shaded}

\begin{longtable}[t]{ll}
\caption{\label{tab:consensus-table}Consensus prior from linear pooling of two experts.}\\
\toprule
Quantity & Value\\
\midrule
Distribution & MIXTURE\\
Components & 2\\
Weights & 0.5, 0.5\\
Mean (SD) & 0.36 (0.1086)\\
\bottomrule
\end{longtable}

Agreement between experts can be quantified by the Bhattacharyya
coefficient (\citeproc{ref-Bhattacharyya1943}{Bhattacharyya 1943}),
which measures distributional overlap on a scale from 0 (no overlap) to
1 (identical distributions). A coefficient close to 1 supports pooling;
a low coefficient indicates the experts' beliefs are largely
non-overlapping, in which case pooling may obscure a genuine
disagreement that should be reported separately.

\subsection{Prior-Data Conflict
Diagnostics}\label{prior-data-conflict-diagnostics}

Prior-data conflict occurs when the observed data is implausible under
the prior predictive distribution (\citeproc{ref-Box1980}{Box 1980}).
\texttt{bayprior} computes five complementary conflict metrics using a
common Normal approximation to the prior and to the observed-data
likelihood (via their means and standard errors), so that the same
diagnostics apply uniformly whatever the prior's distribution family or
the data type:

\begin{itemize}
\tightlist
\item
  \textbf{Prior predictive p-value} (\citeproc{ref-Box1980}{Box 1980}):
  the probability of observing data at least as extreme as the observed
  data under the prior predictive distribution.
\item
  \textbf{S-value} (\citeproc{ref-Greenland2023}{Greenland 2023}): the
  surprisal \(-\log_2(p)\) of the Box p-value, in bits. It is an
  exploratory, continuous companion to the p-value rather than a
  thresholded decision rule, since prior-data conflict checks are
  typically not powered for a binary accept/reject decision.
\item
  \textbf{Surprise index}: the standardised distance between the prior
  mean and the observed data, expressed in prior standard deviation
  units. This differs from the S-value above: it is a distance, not a
  reparametrisation of the p-value, and the two are correlated but
  nonlinearly related.
\item
  \textbf{Information divergence}: the Kullback-Leibler divergence from
  the prior to the normalised likelihood.
\item
  \textbf{Bhattacharyya overlap}: the distributional overlap between the
  prior and normalised likelihood.
\end{itemize}

Writing the prior as \(\pi \sim \mathcal{N}(\mu_\pi, \sigma_\pi^2)\) and
the observed-data likelihood, summarised by its mean and standard error,
as \(\ell \sim \mathcal{N}(\mu_\ell, \sigma_\ell^2)\), all five
diagnostics reduce to closed-form expressions under this common Normal
approximation. The prior predictive distribution of the data summary is
\(\mathcal{N}(\mu_\pi, \sigma_\pi^2 + \sigma_\ell^2)\), giving a
standardised distance
\[z = \frac{\mu_\ell - \mu_\pi}{\sqrt{\sigma_\pi^2 + \sigma_\ell^2}},\]
the two-sided Box p-value \(p = 2\left[1 - \Phi(|z|)\right]\), and the
S-value \(S = -\log_2(p)\). The surprise index uses the prior's own
spread rather than the combined spread of \(z\) above,
\[\text{SI} = \frac{|\mu_\ell - \mu_\pi|}{\sigma_\pi},\] which is why it
is correlated with, but not a monotone reparametrisation of, the Box
p-value. The KL divergence from the prior to the normalised likelihood,
both treated as Normal densities on the same parameter scale, is
\[D_{\mathrm{KL}}(\pi \,\|\, \ell) = \ln\!\left(\frac{\sigma_\ell}{\sigma_\pi}\right)
  + \frac{\sigma_\pi^2 + (\mu_\pi - \mu_\ell)^2}{2\sigma_\ell^2} - \frac{1}{2},\]
and the Bhattacharyya coefficient between the two Normal densities is
\[\mathrm{BC} = \sqrt{\frac{2\sigma_\pi \sigma_\ell}{\sigma_\pi^2 + \sigma_\ell^2}}
  \, \exp\!\left(-\frac{(\mu_\pi - \mu_\ell)^2}{4\left(\sigma_\pi^2 + \sigma_\ell^2\right)}\right).\]
These expressions make explicit that all five diagnostics are functions
of a single standardised quantity, the distance between \(\mu_\pi\) and
\(\mu_\ell\) relative to their spreads, computed after each of the four
supported data types (binary, continuous, Poisson/count, survival) is
mapped to its own approximate mean and standard error, not five
independent pieces of evidence.

\begin{Shaded}
\begin{Highlighting}[]
\NormalTok{interim }\OtherTok{\textless{}{-}} \FunctionTok{list}\NormalTok{(}\AttributeTok{type =} \StringTok{"binary"}\NormalTok{, }\AttributeTok{x =} \DecValTok{18}\NormalTok{, }\AttributeTok{n =} \DecValTok{40}\NormalTok{)}
\NormalTok{cd }\OtherTok{\textless{}{-}} \FunctionTok{prior\_conflict}\NormalTok{(prior, interim)}
\end{Highlighting}
\end{Shaded}

\begin{longtable}[t]{>{\raggedright\arraybackslash}p{3.5cm}>{\raggedright\arraybackslash}p{9cm}}
\caption{\label{tab:cd-table}Prior-data conflict diagnostics for the response rate prior.}\\
\toprule
Diagnostic & Value\\
\midrule
Box's p-value & 0.4319\\
S-value (bits) & 1.21\\
Surprise index & 0.786\\
KL divergence & 0.8761\\
Bhattacharyya overlap & 0.8448\\
\addlinespace
Conflict severity & NONE\\
Recommendation & No evidence of prior-data conflict (Box p = 0.432, S-value = 1.21 bits). The prior appears consistent with the observed data.\\
\bottomrule
\end{longtable}

Four data types are supported for the observed-data side of this
comparison: binary (event count out of a sample size), continuous
(observed mean and SD), Poisson/count (events over exposure), and
survival (events over total follow-up time); each is mapped internally
to an approximate mean and standard error before the
Normal-approximation diagnostics above are applied. This is distinct
from the \emph{exact} conjugate posterior updates (Beta-Binomial,
Normal-Normal, Gamma-Poisson, Gamma-Exponential) that \texttt{bayprior}
uses elsewhere (for example in the sensitivity analysis module below) to
compute posterior summaries.

For co-primary endpoints, multivariate conflict can be assessed using
the Mahalanobis distance (\citeproc{ref-Mahalanobis1936}{Mahalanobis
1936}). The test assumes approximate multivariate normality; for
proportion endpoints such as the illustrative example below,
\texttt{bayprior}'s documentation recommends transforming to the
log-odds scale first. For simplicity, the example here uses raw
proportions directly:

\begin{Shaded}
\begin{Highlighting}[]
\NormalTok{pm   }\OtherTok{\textless{}{-}} \FunctionTok{c}\NormalTok{(}\FloatTok{0.35}\NormalTok{, }\FloatTok{0.60}\NormalTok{)}
\NormalTok{pcov }\OtherTok{\textless{}{-}} \FunctionTok{matrix}\NormalTok{(}\FunctionTok{c}\NormalTok{(}\FloatTok{0.010}\NormalTok{, }\FloatTok{0.003}\NormalTok{, }\FloatTok{0.003}\NormalTok{, }\FloatTok{0.015}\NormalTok{), }\DecValTok{2}\NormalTok{, }\DecValTok{2}\NormalTok{)}
\NormalTok{om   }\OtherTok{\textless{}{-}} \FunctionTok{c}\NormalTok{(}\FloatTok{0.52}\NormalTok{, }\FloatTok{0.58}\NormalTok{)}
\NormalTok{ocov }\OtherTok{\textless{}{-}} \FunctionTok{matrix}\NormalTok{(}\FunctionTok{c}\NormalTok{(}\FloatTok{2e{-}4}\NormalTok{, }\FloatTok{4e{-}5}\NormalTok{, }\FloatTok{4e{-}5}\NormalTok{, }\FloatTok{2e{-}4}\NormalTok{), }\DecValTok{2}\NormalTok{, }\DecValTok{2}\NormalTok{)}

\NormalTok{mv }\OtherTok{\textless{}{-}} \FunctionTok{conflict\_mahalanobis}\NormalTok{(pm, pcov, om, ocov,}
                           \AttributeTok{labels =} \FunctionTok{c}\NormalTok{(}\StringTok{"Response rate"}\NormalTok{, }\StringTok{"OS rate"}\NormalTok{))}
\end{Highlighting}
\end{Shaded}

\begin{longtable}[t]{>{\raggedright\arraybackslash}p{3.5cm}>{\raggedright\arraybackslash}p{9cm}}
\caption{\label{tab:mv-table}Multivariate Mahalanobis conflict check for two co-primary endpoints.}\\
\toprule
Diagnostic & Value\\
\midrule
Mahalanobis D & 1.784\\
Chi-sq p-value (df=2) & 0.2037\\
Conflict flag & FALSE\\
z - Response rate & 1.683\\
z - OS rate & -0.162\\
\addlinespace
Interpretation & No multivariate prior-data conflict detected (Mahalanobis D = 1.784, p = 0.204).\\
\bottomrule
\end{longtable}

\subsection{Sensitivity Analysis}\label{sensitivity-analysis}

The sensitivity module evaluates posterior conclusions across a
hyperparameter grid, answering the regulatory question: does the
treatment benefit conclusion change materially under plausible
alternative priors?

\begin{Shaded}
\begin{Highlighting}[]
\NormalTok{sa }\OtherTok{\textless{}{-}} \FunctionTok{sensitivity\_grid}\NormalTok{(}
  \AttributeTok{prior        =}\NormalTok{ prior,}
  \AttributeTok{data\_summary =}\NormalTok{ interim,}
  \AttributeTok{param\_grid   =} \FunctionTok{list}\NormalTok{(}
    \AttributeTok{alpha =} \FunctionTok{seq}\NormalTok{(}\DecValTok{1}\NormalTok{, }\DecValTok{8}\NormalTok{, }\AttributeTok{by =} \FloatTok{0.5}\NormalTok{),}
    \AttributeTok{beta  =} \FunctionTok{seq}\NormalTok{(}\DecValTok{2}\NormalTok{, }\DecValTok{20}\NormalTok{, }\AttributeTok{by =} \DecValTok{1}\NormalTok{)}
\NormalTok{  ),}
  \AttributeTok{target    =} \FunctionTok{c}\NormalTok{(}\StringTok{"posterior\_mean"}\NormalTok{, }\StringTok{"prob\_efficacy"}\NormalTok{),}
  \AttributeTok{threshold =} \FloatTok{0.30}
\NormalTok{)}
\FunctionTok{plot\_tornado}\NormalTok{(sa)}
\end{Highlighting}
\end{Shaded}

\begin{figure}[H]
\includegraphics[width=1\linewidth]{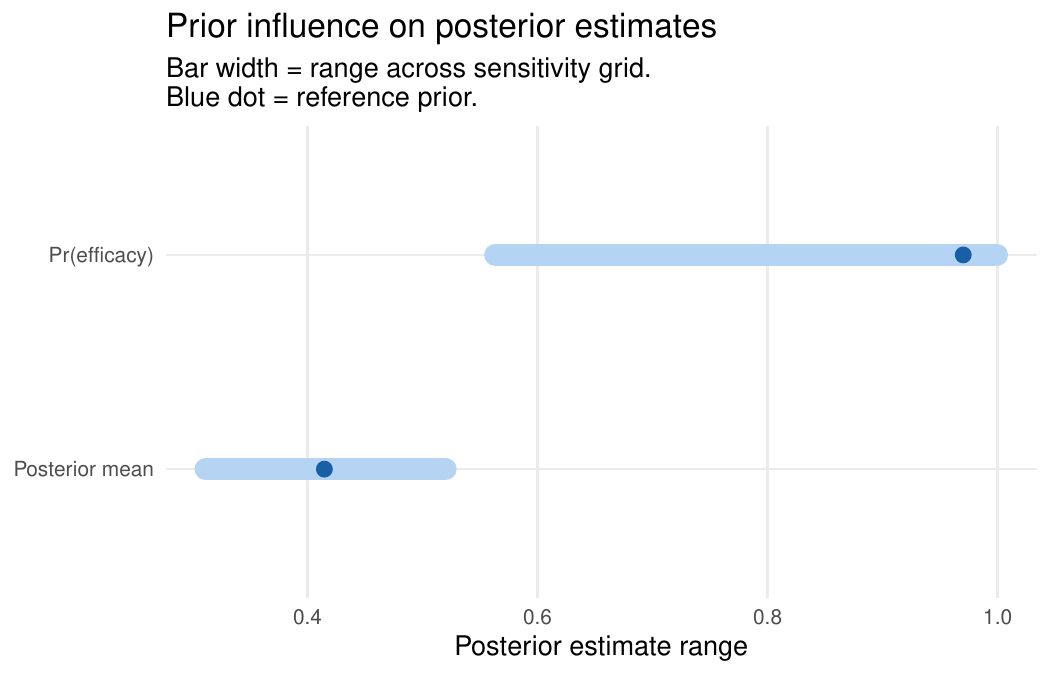} \caption{Tornado plot showing the influence of prior hyperparameters on the posterior mean. Parameters are ranked by their marginal influence.}\label{fig:sensitivity}
\end{figure}

The \texttt{sensitivity\_cri()} function tracks credible interval width
specifically, a quantity directly relevant to the precision of treatment
effect estimates in regulatory submissions.

\subsection{MAP Priors from Historical
Data}\label{map-priors-from-historical-data}

Where the Prior Elicitation and Expert Pooling sections above combine
\emph{expert judgement} into a prior, \texttt{map\_prior()} derives one
directly from \emph{historical trial data} using the
meta-analytic-predictive (MAP) approach of Schmidli et al.
(\citeproc{ref-Schmidli2014}{2014}) and Neuenschwander et al.
(\citeproc{ref-Neuenschwander2010}{2010}). Historical trial point
estimates and standard errors are combined via a random-effects
meta-analysis, with an explicit, outcome-type-specific prior on the
between-trial heterogeneity parameter \(\tau\) following the
weakly-informative recommendations of Roever et al.
(\citeproc{ref-Roever2021}{2021}). Unlike \texttt{robust\_prior()}
(below), which mixes an \emph{already-specified} informative prior with
a vague component for robustness, \texttt{map\_prior()} derives the
informative prior itself from historical data in the first place:

\begin{Shaded}
\begin{Highlighting}[]
\NormalTok{map }\OtherTok{\textless{}{-}} \FunctionTok{map\_prior}\NormalTok{(}
  \AttributeTok{y            =} \FunctionTok{c}\NormalTok{(}\SpecialCharTok{{-}}\FloatTok{0.85}\NormalTok{, }\SpecialCharTok{{-}}\FloatTok{0.62}\NormalTok{, }\SpecialCharTok{{-}}\FloatTok{1.10}\NormalTok{),}
  \AttributeTok{se           =} \FunctionTok{c}\NormalTok{(}\FloatTok{0.25}\NormalTok{, }\FloatTok{0.30}\NormalTok{, }\FloatTok{0.28}\NormalTok{),}
  \AttributeTok{outcome\_type =} \StringTok{"single\_arm\_log\_odds"}\NormalTok{,}
  \AttributeTok{label        =} \StringTok{"Historical control (3 trials)"}
\NormalTok{)}
\FunctionTok{plot\_tau\_posterior}\NormalTok{(map)}
\end{Highlighting}
\end{Shaded}

\begin{figure}[H]
\includegraphics[width=1\linewidth]{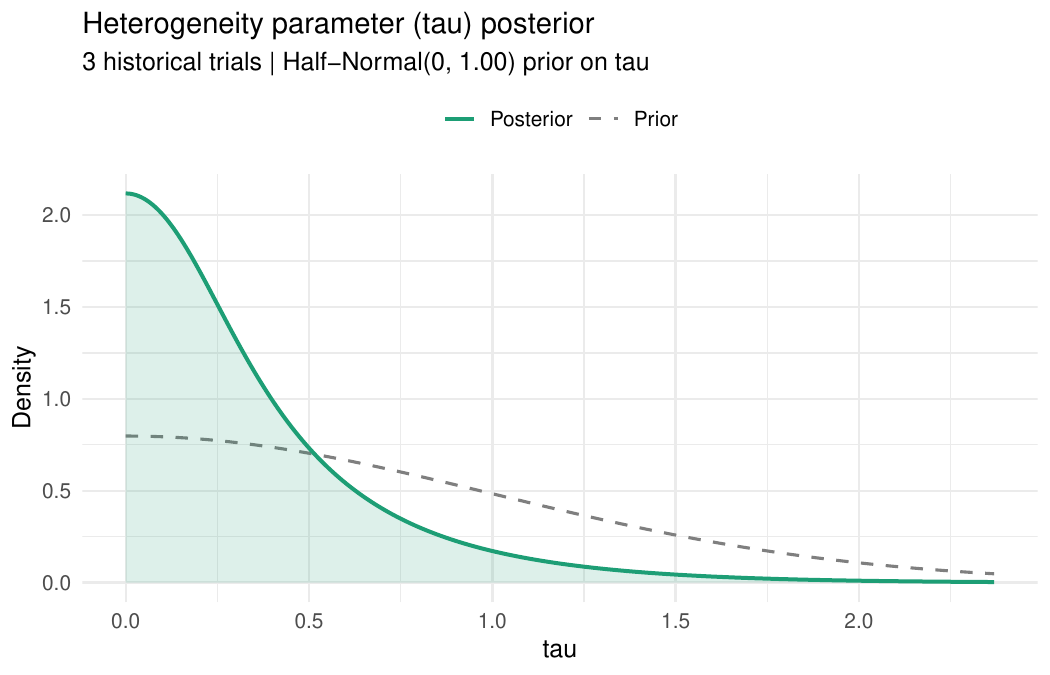} \caption{Posterior density of the between-trial heterogeneity parameter $\tau$, from three historical control-arm trials.}\label{fig:map-prior}
\end{figure}

\begin{longtable}[t]{ll}
\caption{\label{tab:map-table}MAP prior derived from three historical control-arm trials
             (single-arm log-odds scale).}\\
\toprule
Quantity & Value\\
\midrule
Distribution & NORMAL\\
Prior type & map\\
Mean (SD) & -0.8619 (0.3349)\\
95\% CrI & {}[-1.5183, -0.2054]\\
\bottomrule
\end{longtable}

The full \(\tau\) posterior (not just its effect on the meta-analytic
mean) is retained in \texttt{map\$tau\_posterior} and can be exported as
part of \texttt{prior\_report()}, giving reviewers direct visibility
into how much heterogeneity the historical borrowing assumed, rather
than treating it as a fixed nuisance parameter.

\subsection{Robust, Sceptical, and Power
Priors}\label{robust-sceptical-and-power-priors}

\texttt{bayprior} provides three alternative prior types for regulatory
sensitivity analyses.

\textbf{Robust mixture prior} (\citeproc{ref-Schmidli2014}{Schmidli et
al. 2014}): mixes the informative prior with a vague component,
protecting against prior misspecification:

\begin{Shaded}
\begin{Highlighting}[]
\NormalTok{rob }\OtherTok{\textless{}{-}} \FunctionTok{robust\_prior}\NormalTok{(prior, }\AttributeTok{vague\_weight =} \FloatTok{0.20}\NormalTok{)}
\FunctionTok{plot}\NormalTok{(rob)}
\end{Highlighting}
\end{Shaded}

\includegraphics[width=1\linewidth]{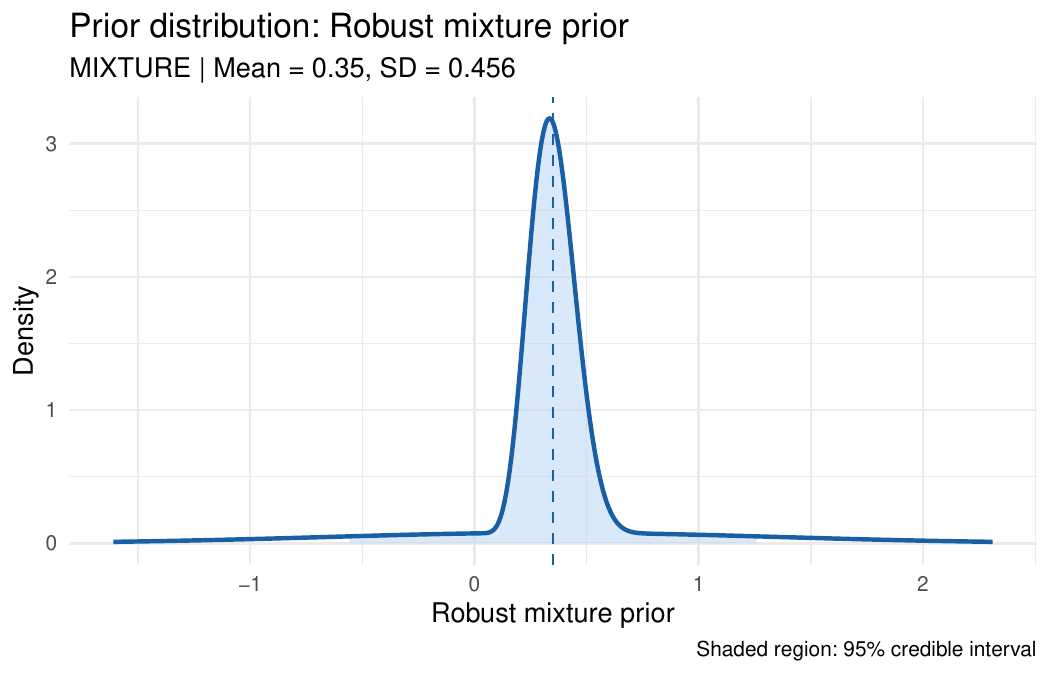}

\textbf{Sceptical prior} (\citeproc{ref-Spiegelhalter1994}{Spiegelhalter
et al. 1994}): centred at the null value of the treatment effect,
representing a conservative regulatory stance:

\begin{Shaded}
\begin{Highlighting}[]
\NormalTok{scep }\OtherTok{\textless{}{-}} \FunctionTok{sceptical\_prior}\NormalTok{(}
  \AttributeTok{null\_value =} \FloatTok{0.20}\NormalTok{,}
  \AttributeTok{family     =} \StringTok{"beta"}\NormalTok{,}
  \AttributeTok{strength   =} \StringTok{"moderate"}\NormalTok{,}
  \AttributeTok{label      =} \StringTok{"Response rate (sceptical)"}
\NormalTok{)}
\end{Highlighting}
\end{Shaded}

\textbf{Power prior} (\citeproc{ref-Ibrahim2000}{Ibrahim and Chen
2000}): down-weights historical data by a factor \(\delta \in (0, 1]\)
calibrated to achieve a target Bayes Factor:

\begin{Shaded}
\begin{Highlighting}[]
\NormalTok{base  }\OtherTok{\textless{}{-}} \FunctionTok{elicit\_beta}\NormalTok{(}\AttributeTok{mean =} \FloatTok{0.50}\NormalTok{, }\AttributeTok{sd =} \FloatTok{0.20}\NormalTok{, }\AttributeTok{method =} \StringTok{"moments"}\NormalTok{,}
                     \AttributeTok{label =} \StringTok{"Response rate"}\NormalTok{)}
\NormalTok{calib }\OtherTok{\textless{}{-}} \FunctionTok{calibrate\_power\_prior}\NormalTok{(}
  \AttributeTok{historical\_data =} \FunctionTok{list}\NormalTok{(}\AttributeTok{type =} \StringTok{"binary"}\NormalTok{, }\AttributeTok{x =} \DecValTok{12}\NormalTok{, }\AttributeTok{n =} \DecValTok{40}\NormalTok{),}
  \AttributeTok{current\_data    =} \FunctionTok{list}\NormalTok{(}\AttributeTok{type =} \StringTok{"binary"}\NormalTok{, }\AttributeTok{x =} \DecValTok{18}\NormalTok{, }\AttributeTok{n =} \DecValTok{50}\NormalTok{),}
  \AttributeTok{base\_prior      =}\NormalTok{ base,}
  \AttributeTok{target\_bf       =} \DecValTok{3}
\NormalTok{)}
\end{Highlighting}
\end{Shaded}

\begin{longtable}[t]{ll}
\caption{\label{tab:calib-table}Power prior calibration via Bayes Factor.}\\
\toprule
Quantity & Value\\
\midrule
Method & bayes\_factor\\
Target BF & 3\\
Optimal delta & 0.05\\
Power prior mean & 0.4448\\
Power prior SD & 0.173\\
\bottomrule
\end{longtable}

\subsection{Regulatory Reporting}\label{regulatory-reporting}

A single call to \texttt{prior\_report()} generates a self-contained
prior justification document in HTML, PDF, or Word format:

\begin{Shaded}
\begin{Highlighting}[]
\FunctionTok{prior\_report}\NormalTok{(}
  \AttributeTok{prior           =}\NormalTok{ prior,}
  \AttributeTok{conflict        =}\NormalTok{ cd,}
  \AttributeTok{sensitivity     =}\NormalTok{ sa,}
  \AttributeTok{robust\_prior    =}\NormalTok{ rob,}
  \AttributeTok{sceptical\_prior =}\NormalTok{ scep,}
  \AttributeTok{output\_format   =} \StringTok{"html"}\NormalTok{,}
  \AttributeTok{trial\_name      =} \StringTok{"TRIAL{-}001"}\NormalTok{,}
  \AttributeTok{sponsor         =} \StringTok{"Example Pharma Ltd"}\NormalTok{,}
  \AttributeTok{author          =} \StringTok{"N.P., Biostatistician"}
\NormalTok{)}
\end{Highlighting}
\end{Shaded}

The generated report includes parameter summary tables, all diagnostic
plots, a compliance checklist aligned with the FDA 2026 draft guidance
(\citeproc{ref-FDA2026}{U.S. Food and Drug Administration 2026}), and a
structured prior justification narrative.

\section{Case study: TRIAL-001}\label{case-study-trial-001}

We demonstrate the \texttt{bayprior} workflow end to end on TRIAL-001, a
synthetic oncology Phase II trial with a binary response rate endpoint.
The target response rate is 30\% (null hypothesis) versus an expected
rate of 40\% (alternative hypothesis). TRIAL-001 is single-arm with no
historical control data, so the power prior and MAP-prior modules are
not demonstrated here; worked examples for both appear in the Package
overview section above.

\subsection{Expert Elicitation and
Pooling}\label{expert-elicitation-and-pooling}

Two clinical experts independently specified Beta priors for the
response rate. Expert 1 (conservative) placed a mean of 0.30 with SD of
0.08; Expert 2 (optimistic) placed a mean of 0.42 with SD of 0.10. Equal
weights were assigned on the basis of comparable seniority.

\begin{Shaded}
\begin{Highlighting}[]
\NormalTok{e1 }\OtherTok{\textless{}{-}} \FunctionTok{elicit\_beta}\NormalTok{(}\AttributeTok{mean =} \FloatTok{0.30}\NormalTok{, }\AttributeTok{sd =} \FloatTok{0.08}\NormalTok{, }\AttributeTok{method =} \StringTok{"moments"}\NormalTok{,}
                  \AttributeTok{label =} \StringTok{"Response rate"}\NormalTok{, }\AttributeTok{expert\_id =} \StringTok{"Expert\_1"}\NormalTok{)}
\NormalTok{e2 }\OtherTok{\textless{}{-}} \FunctionTok{elicit\_beta}\NormalTok{(}\AttributeTok{mean =} \FloatTok{0.42}\NormalTok{, }\AttributeTok{sd =} \FloatTok{0.10}\NormalTok{, }\AttributeTok{method =} \StringTok{"moments"}\NormalTok{,}
                  \AttributeTok{label =} \StringTok{"Response rate"}\NormalTok{, }\AttributeTok{expert\_id =} \StringTok{"Expert\_2"}\NormalTok{)}

\NormalTok{pool }\OtherTok{\textless{}{-}} \FunctionTok{aggregate\_experts}\NormalTok{(}\FunctionTok{list}\NormalTok{(}\AttributeTok{E1 =}\NormalTok{ e1, }\AttributeTok{E2 =}\NormalTok{ e2),}
                          \AttributeTok{weights =} \FunctionTok{c}\NormalTok{(}\FloatTok{0.5}\NormalTok{, }\FloatTok{0.5}\NormalTok{),}
                          \AttributeTok{method  =} \StringTok{"linear"}\NormalTok{)}
\end{Highlighting}
\end{Shaded}

\begin{longtable}[t]{ll}
\caption{\label{tab:pool-table}Consensus prior pooled from two clinical experts (TRIAL-001).}\\
\toprule
Quantity & Value\\
\midrule
Distribution & MIXTURE\\
Components & 2\\
Weights & 0.5, 0.5\\
Mean (SD) & 0.36 (0.1086)\\
\bottomrule
\end{longtable}

The two expert distributions overlap substantially (means 0.30 and 0.42
with comparable SDs of 0.08 and 0.10), supporting the use of
equal-weight linear pooling. The consensus prior had a mean of 0.36 and
SD of 0.109.

\subsection{Conflict Diagnostics at
Interim}\label{conflict-diagnostics-at-interim}

At the pre-specified interim analysis, 18 responses were observed in 40
patients (observed rate: 45\%). Following common practice for Box's
prior predictive check, a significance threshold of \(\alpha = 0.05\)
was pre-specified: a Box p-value below this threshold would indicate
evidence of prior-data conflict warranting reassessment of the prior.
This is the default threshold in \texttt{prior\_conflict()}.

\begin{Shaded}
\begin{Highlighting}[]
\NormalTok{interim\_data }\OtherTok{\textless{}{-}} \FunctionTok{list}\NormalTok{(}\AttributeTok{type =} \StringTok{"binary"}\NormalTok{, }\AttributeTok{x =} \DecValTok{18}\NormalTok{, }\AttributeTok{n =} \DecValTok{40}\NormalTok{)}
\NormalTok{cd\_trial }\OtherTok{\textless{}{-}} \FunctionTok{prior\_conflict}\NormalTok{(pool, interim\_data, }\AttributeTok{alpha =} \FloatTok{0.05}\NormalTok{)}
\end{Highlighting}
\end{Shaded}

\begin{longtable}[t]{>{\raggedright\arraybackslash}p{3.5cm}>{\raggedright\arraybackslash}p{9cm}}
\caption{\label{tab:cdtrial-table}Prior-data conflict diagnostics at the TRIAL-001 interim analysis.}\\
\toprule
Diagnostic & Value\\
\midrule
Box's p-value & 0.5022\\
S-value (bits) & 0.99\\
Surprise index & 0.6711\\
KL divergence & 0.7853\\
Bhattacharyya overlap & 0.8709\\
\addlinespace
Conflict severity & NONE\\
Recommendation & No evidence of prior-data conflict (Box p = 0.502, S-value = 0.99 bits). The prior appears consistent with the observed data.\\
\bottomrule
\end{longtable}

The Box p-value was above the pre-specified 0.05 threshold, indicating
no evidence of prior-data conflict. The posterior distribution updated
the consensus prior towards the observed data as expected.

\subsection{Enthusiastic and Sceptical
Priors}\label{enthusiastic-and-sceptical-priors}

A common regulatory sensitivity practice, following Spiegelhalter et al.
(1994), is to present conclusions under both an enthusiastic prior
(favouring treatment benefit) and a sceptical prior (centred at the
null), alongside the consensus prior already assessed above:

\begin{Shaded}
\begin{Highlighting}[]
\NormalTok{enth\_trial }\OtherTok{\textless{}{-}} \FunctionTok{elicit\_beta}\NormalTok{(}\AttributeTok{mean =} \FloatTok{0.45}\NormalTok{, }\AttributeTok{sd =} \FloatTok{0.08}\NormalTok{, }\AttributeTok{method =} \StringTok{"moments"}\NormalTok{,}
                          \AttributeTok{label =} \StringTok{"Response rate (enthusiastic)"}\NormalTok{)}
\NormalTok{scep\_trial }\OtherTok{\textless{}{-}} \FunctionTok{sceptical\_prior}\NormalTok{(}\AttributeTok{null\_value =} \FloatTok{0.20}\NormalTok{, }\AttributeTok{family =} \StringTok{"beta"}\NormalTok{,}
                              \AttributeTok{strength =} \StringTok{"moderate"}\NormalTok{,}
                              \AttributeTok{label =} \StringTok{"Response rate (sceptical)"}\NormalTok{)}

\NormalTok{cd\_enth }\OtherTok{\textless{}{-}} \FunctionTok{prior\_conflict}\NormalTok{(enth\_trial, interim\_data, }\AttributeTok{alpha =} \FloatTok{0.05}\NormalTok{)}
\NormalTok{cd\_scep }\OtherTok{\textless{}{-}} \FunctionTok{prior\_conflict}\NormalTok{(scep\_trial, interim\_data, }\AttributeTok{alpha =} \FloatTok{0.05}\NormalTok{)}

\FunctionTok{cat}\NormalTok{(}\StringTok{"Conflict severity (consensus):    "}\NormalTok{, cd\_trial}\SpecialCharTok{$}\NormalTok{conflict\_severity, }\StringTok{"}\SpecialCharTok{\textbackslash{}n}\StringTok{"}\NormalTok{)}
\end{Highlighting}
\end{Shaded}

\begin{verbatim}
#> Conflict severity (consensus):     none
\end{verbatim}

\begin{Shaded}
\begin{Highlighting}[]
\FunctionTok{cat}\NormalTok{(}\StringTok{"Conflict severity (enthusiastic): "}\NormalTok{, cd\_enth}\SpecialCharTok{$}\NormalTok{conflict\_severity, }\StringTok{"}\SpecialCharTok{\textbackslash{}n}\StringTok{"}\NormalTok{)}
\end{Highlighting}
\end{Shaded}

\begin{verbatim}
#> Conflict severity (enthusiastic):  none
\end{verbatim}

\begin{Shaded}
\begin{Highlighting}[]
\FunctionTok{cat}\NormalTok{(}\StringTok{"Conflict severity (sceptical):    "}\NormalTok{, cd\_scep}\SpecialCharTok{$}\NormalTok{conflict\_severity, }\StringTok{"}\SpecialCharTok{\textbackslash{}n}\StringTok{"}\NormalTok{)}
\end{Highlighting}
\end{Shaded}

\begin{verbatim}
#> Conflict severity (sceptical):     mild
\end{verbatim}

The consensus and enthusiastic priors showed no evidence of prior-data
conflict (Box p-values of 0.5 and 1); the sceptical prior showed mild
conflict (Box p = 0.026). This is the expected diagnostic outcome for a
prior deliberately centred at the null response rate (20\%) confronted
with a substantially higher observed interim rate (45\%): the sceptical
prior is a stress test by design, and the diagnostic correctly flags
that it is in tension with the data, rather than indicating a problem
with the trial's conclusion. The consensus and enthusiastic priors,
which reflect the experts' actual clinical belief rather than a
deliberately adversarial position, are not flagged.

\subsection{Sensitivity Analysis}\label{sensitivity-analysis-1}

Beyond comparing discrete alternative priors, the sensitivity module
evaluates how the posterior conclusion changes continuously across a
hyperparameter grid. Because the grid is defined over the
hyperparameters of a single distribution family,
\texttt{sensitivity\_grid()} moment-matches a mixture (pooled) prior's
actual blended mean and SD to a working Beta prior before gridding,
rather than analysing a single expert's prior in isolation; a message
reports the working prior used (here, mean \(\approx\) 0.36, SD
\(\approx\) 0.11, reflecting both experts):

\begin{Shaded}
\begin{Highlighting}[]
\NormalTok{sa\_trial }\OtherTok{\textless{}{-}} \FunctionTok{sensitivity\_grid}\NormalTok{(}
  \AttributeTok{prior        =}\NormalTok{ pool,}
  \AttributeTok{data\_summary =}\NormalTok{ interim\_data,}
  \AttributeTok{param\_grid   =} \FunctionTok{list}\NormalTok{(}
    \AttributeTok{alpha =} \FunctionTok{seq}\NormalTok{(}\DecValTok{3}\NormalTok{, }\DecValTok{12}\NormalTok{, }\AttributeTok{by =} \DecValTok{1}\NormalTok{),}
    \AttributeTok{beta  =} \FunctionTok{seq}\NormalTok{(}\DecValTok{6}\NormalTok{, }\DecValTok{20}\NormalTok{, }\AttributeTok{by =} \DecValTok{1}\NormalTok{)}
\NormalTok{  ),}
  \AttributeTok{target    =} \FunctionTok{c}\NormalTok{(}\StringTok{"posterior\_mean"}\NormalTok{, }\StringTok{"prob\_efficacy"}\NormalTok{),}
  \AttributeTok{threshold =} \FloatTok{0.30}
\NormalTok{)}
\FunctionTok{plot\_tornado}\NormalTok{(sa\_trial)}
\end{Highlighting}
\end{Shaded}

\begin{figure}[H]
\includegraphics[width=1\linewidth]{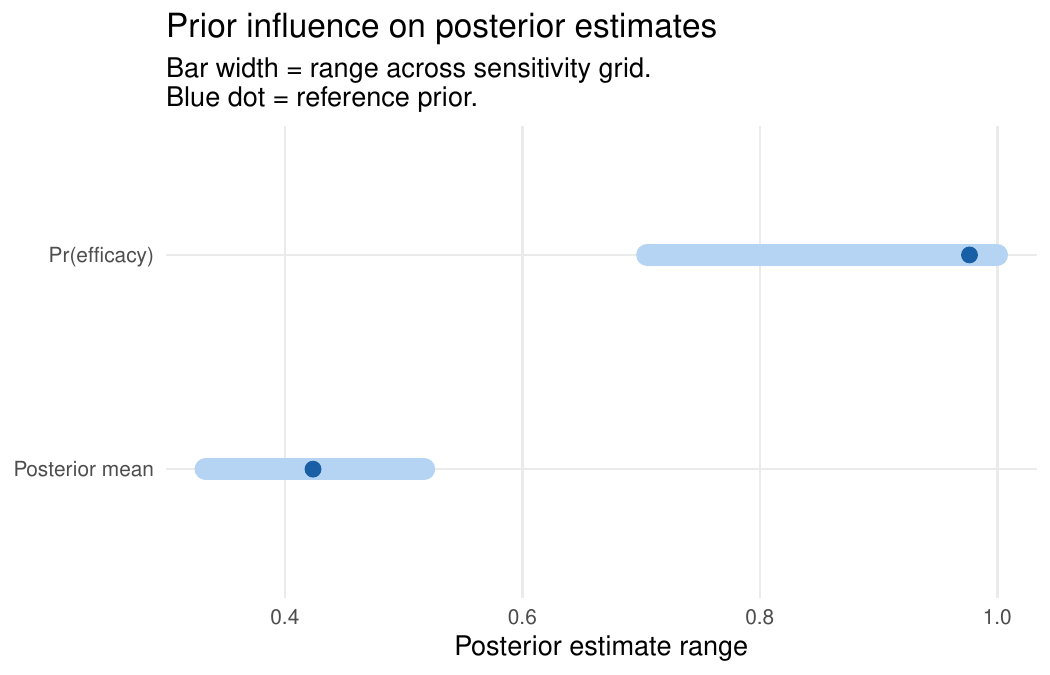} \caption{Tornado plot for TRIAL-001: influence of the moment-matched pooled prior's Beta hyperparameters on the posterior mean, given the interim data.}\label{fig:trial-sensitivity}
\end{figure}

Across the explored hyperparameter range, the posterior mean varies from
0.33 to 0.52 and the probability of exceeding the 30\% efficacy
threshold varies from 0.71 to 1. Neither quantity is pinned to a single
value, but the trial's qualitative conclusion does not reverse anywhere
in the grid: the posterior mean remains above the 30\% null in every
cell, and the probability of efficacy never approaches a level that
would change a regulatory decision built on this threshold. The trial's
conclusion is therefore not driven by fine details of the pooled prior's
specification, even though the exact posterior summaries do move with
the hyperparameters.

\subsection{Robust Mixture Prior}\label{robust-mixture-prior}

As a further regulatory sensitivity check, a robust mixture prior mixes
the consensus prior with a vague component, bounding the influence any
single prior specification can have on the posterior:

\begin{Shaded}
\begin{Highlighting}[]
\NormalTok{rob\_trial }\OtherTok{\textless{}{-}} \FunctionTok{robust\_prior}\NormalTok{(pool, }\AttributeTok{vague\_weight =} \FloatTok{0.20}\NormalTok{)}
\NormalTok{cd\_rob    }\OtherTok{\textless{}{-}} \FunctionTok{prior\_conflict}\NormalTok{(rob\_trial, interim\_data, }\AttributeTok{alpha =} \FloatTok{0.05}\NormalTok{)}

\FunctionTok{cat}\NormalTok{(}\StringTok{"Conflict severity (robust mixture):"}\NormalTok{, cd\_rob}\SpecialCharTok{$}\NormalTok{conflict\_severity, }\StringTok{"}\SpecialCharTok{\textbackslash{}n}\StringTok{"}\NormalTok{)}
\end{Highlighting}
\end{Shaded}

\begin{verbatim}
#> Conflict severity (robust mixture): none
\end{verbatim}

The robust mixture prior showed the same conflict conclusion (no
evidence of conflict) as the consensus and enthusiastic priors above. It
does not match the sceptical prior's mild-conflict flag, nor should it:
the sceptical prior is a deliberately adversarial stress test, while the
robust mixture is a bounded-influence version of the experts' actual
consensus belief, so the two are expected to diverge.

\subsection{Regulatory Report}\label{regulatory-report}

The full set of TRIAL-001 analyses (the pooled prior, conflict
diagnostics, sensitivity analysis, and robust and sceptical
alternatives) can be assembled into a single submission-ready document
with one function call:

\begin{Shaded}
\begin{Highlighting}[]
\FunctionTok{prior\_report}\NormalTok{(}
  \AttributeTok{prior           =}\NormalTok{ pool,}
  \AttributeTok{conflict        =}\NormalTok{ cd\_trial,}
  \AttributeTok{sensitivity     =}\NormalTok{ sa\_trial,}
  \AttributeTok{robust\_prior    =}\NormalTok{ rob\_trial,}
  \AttributeTok{sceptical\_prior =}\NormalTok{ scep\_trial,}
  \AttributeTok{output\_format   =} \StringTok{"html"}\NormalTok{,}
  \AttributeTok{trial\_name      =} \StringTok{"TRIAL{-}001"}\NormalTok{,}
  \AttributeTok{sponsor         =} \StringTok{"Example Pharma Ltd"}\NormalTok{,}
  \AttributeTok{author          =} \StringTok{"N.P., Biostatistician"}
\NormalTok{)}
\end{Highlighting}
\end{Shaded}

\section{Discussion}\label{discussion}

\texttt{bayprior} provides a structured workflow for prior justification
that addresses a practical gap in the Bayesian clinical trial toolkit.
It is intended for use alongside, not as a replacement for, established
posterior-computation packages such as \texttt{rstanarm}
(\citeproc{ref-rstanarm}{Goodrich et al. 2024}) and \texttt{brms}
(\citeproc{ref-Buerkner2017}{Bürkner 2017}): bayprior handles the prior
specification phase, and these packages handle model fitting. bayprior's
fitted hyperparameters are ordinary distribution parameters rather than
an opaque object (the \texttt{prior\$params\$alpha}/\texttt{beta} shown
in Table 2 above, or a \texttt{map\_prior()} fit's Normal mean and SD on
the log-odds scale), and for the latter this is exactly the Normal
intercept-prior parameterisation both downstream packages expect for a
logistic-regression model, requiring no distributional-family
conversion. An example of both handoffs, including the syntax each
package's own prior-specification API actually requires (which is easy
to get wrong), is provided in the package's
\texttt{vignette("using-with-rstanarm-brms")}. \texttt{rstanarm} and
\texttt{brms} are listed only as optional (\texttt{Suggests})
dependencies for this demonstration; neither is required to use
\texttt{bayprior} itself.

\subsection{Regulatory Alignment}\label{regulatory-alignment}

The FDA's 2026 draft guidance on Bayesian methods
(\citeproc{ref-FDA2026}{U.S. Food and Drug Administration 2026}) asks
sponsors relying on a Bayesian analysis to provide detailed support for
the proposed prior distribution and any external information borrowing,
to quantify the prior's influence (for example via an effective sample
size), to evaluate operating characteristics (including sensitivity to
the prior specification and to prior-data conflict, typically via
simulation) and to document the analysis in sufficient detail for the
FDA to understand and, where feasible, reproduce it. \texttt{bayprior}
supports each of these expectations directly: structured elicitation and
pooling provide a documented, reproducible derivation of the prior;
\texttt{map\_prior()} gives a closed-form, auditable account of how much
external/historical information was borrowed and with what assumed
heterogeneity; the sensitivity, robust, sceptical, and power prior
modules provide alternative specifications for demonstrating robustness;
and the regulatory reporting module assembles this into a single
submission-ready document.

Prior-data conflict and the influence of external information are also a
recurring theme in EMA guidance on incorporating historical or external
data. The EMA reflection paper on extrapolation in paediatric drug
development (\citeproc{ref-EMA2018}{European Medicines Agency 2018}),
for instance, requires sponsors to quantify how much information a
Bayesian prior contributes relative to the data generated in the target
population and to investigate the Type I error properties of the
resulting analysis. The conflict diagnostics in \texttt{bayprior} (the
Box predictive p-value, S-value, surprise index, information divergence,
and Bhattacharyya overlap) provide a closed-form way to address this for
conjugate prior families, without requiring a fitted model.

\subsection{Shiny Application}\label{shiny-application}

The bundled Shiny application mirrors the package's analytical workflow
as separate panels, allowing a biostatistician without R programming
experience to perform the same elicitation, pooling, MAP-prior
derivation, conflict, sensitivity, robust prior, and reporting workflow
interactively. Each panel produces the same plots and summary statistics
as the corresponding R function, and the final regulatory report can be
downloaded directly from the application. The application is hosted at
\url{https://ndohpenn-bayprior.share.connect.posit.cloud} and can be
launched locally via \texttt{run\_app()}.

\subsection{Limitations}\label{limitations}

Several limitations apply to the current release (v0.4.0). The
multivariate Mahalanobis check is currently limited to bivariate (k = 2)
endpoints; extension to k \(\geq\) 3 co-primary endpoints is planned for
a future release. Survival power priors require numerical integration
over the partial likelihood and are not yet implemented; calibrated
power priors are currently available for binary, continuous, and
Poisson/count data, using Beta, Normal, Gamma, or Lognormal prior
families as appropriate. The robust mixture prior uses a Normal vague
component regardless of the informative prior family; for Beta, Gamma,
and other non-Normal informative priors, the resulting mixture density
is computed numerically rather than analytically, which is accurate in
the body of the distribution but may be less reliable in the extreme
tails. Sensitivity grids over mixture priors moment-match the pooled
mean and SD to a working prior in the dominant component's family (Beta,
Normal, Gamma, or Log-Normal); families without a mean/SD
parameterisation (Exponential, Weibull) fall back to the dominant
component alone, with a warning. Finally, the conflict diagnostics
themselves rest by default on a Normal approximation to the prior and to
the data likelihood (see the Prior-Data Conflict Diagnostics section
above). \texttt{prior\_conflict(...,\ exact\ =\ TRUE)} replaces this
with the exact prior predictive distribution: Beta-Binomial for a Beta
prior against binary data, and Negative Binomial (via Gamma-Poisson
conjugacy) for a Gamma prior against count or survival data (the two
share one exact check, since both reduce to the same events/exposure
model). This eliminates the approximation entirely for these common
cases; the Normal-Normal case (a Normal prior against continuous data)
is already exact under the default. For prior families or data types
without a closed-form exact check (for example a Weibull-elicited
prior), the Normal approximation remains the only option, and this paper
does not separately quantify its accuracy there, nor the
repeated-sampling operating characteristics (for example, the false
conflict-flag rate under the null) of the resulting diagnostics more
generally.

Future development includes survival power priors and k \(\geq\) 3
endpoint Mahalanobis conflict. A companion paper in Pharmaceutical
Statistics (in preparation) provides a methodological treatment of the
conflict diagnostic framework, including this operating-characteristics
assessment, and a regulatory alignment assessment against the FDA 2026
draft guidance.

\section{Summary}\label{summary}

\texttt{bayprior} provides an integrated R tool for the complete prior
justification workflow in Bayesian clinical trials, combining structured
expert elicitation, opinion pooling, MAP priors from historical trial
data, conjugate conflict diagnostics, hyperparameter sensitivity
analysis, robust prior alternatives, and regulatory reporting in a
single package. The package is available on CRAN:

\begin{Shaded}
\begin{Highlighting}[]
\FunctionTok{install.packages}\NormalTok{(}\StringTok{"bayprior"}\NormalTok{)}
\end{Highlighting}
\end{Shaded}

The full documentation and vignettes are available at
\url{https://ndohpenngit.github.io/bayprior/}. A live Shiny application
is hosted at \url{https://ndohpenn-bayprior.share.connect.posit.cloud}.

To cite \texttt{bayprior} in publications, use:

\begin{verbatim}
Penn N (2026). bayprior: Bayesian Prior Elicitation, Diagnostics, and
Regulatory Reporting. R package version 0.4.0,
<https://CRAN.R-project.org/package=bayprior>.
\end{verbatim}

\section*{References}\label{bibliography}
\addcontentsline{toc}{section}{References}

\protect\phantomsection\label{refs}

\end{document}